\documentstyle[epsf,12pt]{article}

\begin{document}

\begin{titlepage}
\begin{flushright}
OSU-HEP-01-09\\
BA-01-49\\
November 2001\\
\end{flushright}
\vskip 2cm
\begin{center}
{\Large\bf
Bimaximal Neutrino Mixings from\\}
\vspace*{0.1in}
{\Large\bf Lopsided Mass Matrices 
}
\vskip 1cm
{\large\bf
K.S.\ Babu$\,{}^1$ and S.M. Barr$\,{}^{2}$} \\
\vskip 0.5cm
{\it ${}^1\,$Department of Physics, Oklahoma State University\\
Stillwater, OK~~74078, USA\\ [0.1truecm]
${}^2\,$Bartol Research Institute, University of Delaware\\
Newark, DE~~ 19716, USA\\[0.1truecm]
}

\end{center}
\vskip .5cm

\begin{abstract}

Current solar and atmospheric neutrino oscillation data seem
to favor a bimaximal pattern for neutrino mixings where the matrix elements
$U_{e2}$ and $U_{\mu 3}$ are of order one, while $U_{e3}$ is much smaller.
We show that such a pattern can be obtained quite easily in
theories with ``lopsided'' mass matrices for the charged leptons
and the down type quarks.  A relation connecting the solar and atmospheric
neutrino mixing angles is derived, $\tan^2\theta_{atm} \simeq 1+
\tan^2\theta_{sol}$, which predicts $\sin^22\theta_{atm} \simeq 0.97$
corresponding to the best fit LMA solution for solar neutrinos.  Predictive
schemes in $SO(10)$ realizing these ideas are presented.  A new class of
$SO(10)$ models with lopsided mass matrices is found which makes use of
an adjoint VEV along the $I_{3R}$ direction, rather than the traditional
$B-L$ direction.

\end{abstract}

\end{titlepage}

\noindent {\large \bf \S 1. Introduction}
\vspace*{0.1in}

Recent data seems somewhat to favor either the LMA (large mixing angle MSW)
or the LOW solution to the solar neutrino problem over the small mixing
angle solution \cite{solar,review}.
Taken together with the atmospheric neutrino results \cite{atm} and the
CHOOZ reactor experiment \cite{chooz}, this would imply a so-called
``bimaximal'' pattern of mixing, with $U_{e2}$ and $U_{\mu 3}$ large and
$U_{e3}$ small ($U$ is the leptonic mixing matrix) \cite{bimax}. Another possibly
significant feature of the data is
that the atmospheric neutrino mixing angle is not merely large, but seems
to be nearly maximal. The best-fit value at present is $\sin^2 2 \theta_{atm}
\simeq 1.0$ \cite{atm}

In this paper we make several points relevant to these observations.
\begin{enumerate}

\item The so-called ``lopsided" models
\cite{bb,ab1,abb,syilr,ab2,lopsided} provide a very simple way of
accounting for
the bimaximal pattern of mixing, and in particular have no difficulty in
obtaining the LMA solar solution, unlike certain other kinds of bimaximal
schemes.

\item By combining the idea of lopsided mass matrices
with a nonabelian flavor symmetry one can explain in a simple way the
near maximality of $\theta_{atm}$. In the simplest case one
obtains the relation $\tan^2 \theta_{atm} = 1 + \tan^2 \theta_{sol}$.
In the limit of small solar angle this gives maximal atmospheric angle;
while for the best-fit LMA value \cite{bgp} of $\tan^2 \theta_{sol} \cong 0.4$,
it gives $\sin^2 2 \theta_{atm} = 0.97$.

\item The lopsided bimaximal
idea can be straightforwardly implemented in the context of $SO(10)$,
and in that case a prediction relating quark masses and mixings to the
atmospheric neutrino mixing angle arises.

\item A new class of predictive $SO(10)$ models for quark and lepton masses is
found which makes use of an adjoint VEV along
the $I_{3R}$ direction.  Bimaximal mixing pattern for neutrinos
can be obtained easily in this class of models, along with several predictions
relating the charged fermion masses and mixings.  This provides a new way
of looking at quark and lepton masses in $SO(10)$, different from the traditional
way where the VEV of the adjoint points along the $B-L$ direction.

\end{enumerate}

\vspace*{0.1in}
\noindent {\large \bf \S 2.  Bimaximal mixing}
\vspace*{0.1in}

Imagine that the leptonic mixing angles
come primarily from the diagonalization of the {\it charged} lepton mass
matrix $L$, which has the following ``lopsided" form:

\begin{equation}
L = \left( \begin{array}{ccc} - & - & - \\
- & - & \epsilon \\ \rho' & \rho & 1 \end{array} \right) m_D.
\end{equation}

\noindent
Here $\rho' \sim \rho \sim 1$, whereas $\epsilon \ll 1$. The dashes represent
elements that are small compared to the ones shown. The convention
being used is that the left-handed lepton fields multiply the mass matrix from
the right. The diagonalization of this matrix can be done in stages, the
first stage being to rotate in the space of $\ell_2^-$ and $\ell_1^-$ by
an angle which we will call $\theta_s$, satisfying $\tan \theta_s
= \rho'/\rho$. This brings the matrix to the form

\begin{equation}
L' = \left( \begin{array}{ccc} - & - & - \\
- & - & \epsilon \\ 0 & \sigma & 1 \end{array} \right) m_D~,
\end{equation}

\noindent
where $\sigma \equiv \sqrt{\rho^{\prime 2} + \rho^2}$. (Note that all
parameters shown in $L$ can be made real by field redefinitions.)
The next stage
is to rotate in the space of $\ell_3^-$ and the new $\ell_2^-$ by an angle
which we will call $\theta_a$, satisfying $\tan \theta_a
= \sigma$. This brings the matrix to the form

\begin{equation}
L^{\prime \prime} = \left( \begin{array}{ccc} - & - & - \\
- & \frac{- \sigma}{\sqrt{\sigma^2 + 1}} \epsilon &
\frac{1}{\sqrt{\sigma^2 + 1}} \epsilon \\
0 & 0 & \sqrt{\sigma^2 + 1}  \end{array} \right) m_D.
\end{equation}

The rotations needed to complete the diagonalization involve only small
rotations of the left-handed leptons, and we will therefore neglect them.
(An important point is that the (2,1) and (2,2) elements of $L$ were assumed small
compared to $\epsilon$. Otherwise, there would still be required a large
rotation in the 1-2 plane to diagonalize $L^{\prime \prime}$, and that would
induce a large $U_{e3}$.)
The unitary matrix $U_L$ required to diagonalize $L^{\dag} L$ is thus
approximately

\begin{equation}
U_L^{\dag}  \cong
\left( \begin{array}{ccc} 1 & 0 & 0 \\ 0 & \cos \theta_a &
- \sin \theta_a \\ 0 & \sin \theta_a & \cos \theta_a
\end{array} \right)
\left( \begin{array}{ccc}
\cos \theta_s & - \sin \theta_s & 0 \\ \sin \theta_s &
\cos \theta_s & 0 \\ 0 & 0 & 1 \end{array} \right)
= \left( \begin{array}{ccc}
c_s & s_s & 0 \\ c_a s_s & c_a s_s & - s_a \\
s_a s_s & s_a c_s & c_a \end{array} \right).
\end{equation}

\noindent
The full leptonic mixing matrix is given by $U_{MNS} =
U^{\dag}_L U_{\nu}$, where
$U_{\nu}$ is the unitary matrix required to diagonalize the neutrino
mass matrix. However, since we are assuming $U_{\nu} \cong I$,
$U_{MNS}$ is given approximately by the matrix in Eq. (4), which
has the bimaximal mixing pattern, with $U_{e2}$ and $U_{\mu 3}$
both of order unity and $U_{e3}$ small. A very important point is
that no constraint whatsoever has had to be placed on the neutrino
{\it masses}. The questions of neutrino mixing and neutrino mass are
completely decoupled in this scenario. This means, in particular, that
there is no difficulty in obtaining the neutrino mass ratios
appropriate to any of the large angle solar solutions, LMA, LOW, and VAC.

In many published models the bimaximal mixing comes from the diagonalization
of the neutrino mass matrix \cite{bd}. It is instructive to compare the present
idea to some of these other approaches. Consider the following three
forms of $M_{\nu}$, the light neutrino mass matrix obtained after seesaw
diagonalization.

\begin{equation}
M^A_{\nu} \sim \left( \begin{array}{ccc}
- & - & - \\ - & \sigma^2 & \sigma \\ - & \sigma & 1 \end{array}
\right)m_{\nu}, \;\;\;
M^B_{\nu} \sim \left( \begin{array}{ccc}
\rho^{\prime 2} & \rho \rho' & \rho' \\ \rho \rho' & \rho^2 & \rho \\
\rho' & \rho & 1 \end{array} \right)m_{\nu}, \;\;\;
M^C_{\nu} \sim \left( \begin{array}{ccc}
- & 1 & \sigma \\ 1 & - & - \\ \sigma & - & - \end{array} \right)m_{\nu}.
\end{equation}

\noindent
As before, the dashes indicate elements smaller than the ones explicitly
shown, and $\rho$, $\rho'$ and $\sigma$ are assumed to be of order unity.

In matrix $M_{\nu}^A$, a large rotation angle, satisfying
$\tan \theta_a \cong \sigma$, is required to diagonalize the 2-3 block.
This produces a large atmospheric neutrino mixing angle. The magnitude of
the solar neutrino mixing angle depends on the magnitude of the small
elements in $M_{\nu}^A$, and may also be large.
Because of the approximately ``factorized" or rank-1
form of the 2-3 block of this
matrix, there is only one large mass eigenvalue, so that the desired
hierarchy $m_1, m_2 \ll m_3$ results. Matrix $M_{\nu}^A$ can thus give
a satisfactory
bimaximal mixing. However, there is a price to be paid for this: in order
for both the mixing angles and the neutrino masses to come out right a certain
precise relationship had to be assumed to exist among the elements of
$M_{\nu}$ --- namely, the approximately factorized or rank-1 structure of
the 2-3 block. Moreover, for the solar neutrino mixing angle to be large,
further assumptions have to be made about the small elements of $M_{\nu}^A$.

Matrix $M_{\nu}^B$ has the apparent advantage over matrix $M_{\nu}^A$
that both the atmospheric and the solar neutrino mixing angles automatically
come out to be order one. However, it does not give a realistic bimaximal
scheme, since $U_{e3}$ is of order one rather than small. The reason is the
following. To diagonalize $M_{\nu}^B$, requires first rotating in the
1-2 plane by an angle with $\tan \theta_s = \rho'/\rho$, and then in the 2-3
plane by an angle with $\tan \theta_a = \sigma \equiv \sqrt{\rho^{\prime 2} +
\rho^2}$. This is the same as what was required to diagonalize the charged
lepton mass matrix in Eq. (1). However, because one is diagonalizing
the {\it neutrino} mass matrix in this case, the resulting MNS matrix
is the {\it adjoint} of what was obtained in Eq. (4). (Recall that $U_{MNS}
= U^{\dag}_L U_{\nu}$.) Thus, here $U_{e3} = \sin \theta_a \sin \theta_s$.
Moreover, as in
the previous example, matrix $M_{\nu}^B$ requires a form in which the elements
are in a special precise relationship to each other.

Matrix $M_{\nu}^C$ is the typical ``inverted hierarchy" form \cite{ih},
with an approximate $L_e-L_\mu-L_\tau$ symmetry and
automatically gives bimaximal mixing, with $U_{e3}$ very small.
This can be seen as follows. The matrix $M_{\nu}^C$ can be diagonalized
in stages, as in the other examples. In this case, however, the first stage
is to rotate in the 2-3 plane by an angle $\theta_a$ such that
$\tan \theta_a = \sigma$. This brings the matrix to a ``pseudo-Dirac"
form with large and equal (1,2) and (2,1) elements and all other elements small.
The next stage is a rotation by $\pi/4$ in the 1-2 plane. Thus, $U_{\nu}$
has the form given in Eq. (4) with $\theta_s \cong \pi/4$. The fact that the
solar neutrino angle typically comes out very close to maximal is
certainly acceptable for the LOW and VAC solutions, but may not be acceptable
for the LMA solution if it turns out that the LMA fits
require $\tan^2 \theta_{sol}$ to be significantly smaller than one.
At the moment, the best-fit LMA value is $\tan^2 \theta_{sol} \sim 0.4$,
but maximal mixing is within the 99\% confidence level contours given in
\cite{bgp}.

One sees from these comparisons that obtaining bimaximal mixing from the
diagonalization of the charged lepton mass matrix simplifies the problem by
neatly separating the questions of neutrino mass and neutrino mixing.

\vspace{0.1in}
\noindent {\large \bf \S 3. Nearly maximal atmospheric neutrino mixing}
\vspace*{0.1in}

At present the best fit to the atmospheric neutrino angle is $\sin^2 2
\theta_{atm} \simeq 1.0$.
This is difficult to obtain as a prediction from models. Several types
of model, indeed, predict that the {\it solar} angle should be very close
to maximal --- for instance, the ``inverted hierarchy" models \cite{ih}
just described and ``flavor democracy" models \cite{fd}.
But maximal {\it atmospheric} neutrino mixing is
much harder to achieve. The reason is simple. The most obvious way to
get nearly maximal mixing of two neutrino flavors is by a pseudo-Dirac
form of the mass matrix: $\left( \begin{array}{cc} \delta & 1 \\
1 & \delta' \end{array} \right)$, with $\delta, \delta' \ll 1$.
This form also gives nearly degenerate
neutrinos. Therefore, if such a form is assumed for the 1-2 block of
$M_{\nu}$, to give maximal solar neutrino mixing angle, it also typically
gives $\Delta m^2_{sol} \ll \Delta m^2_{atm}$, as desired. However, if the
2-3 block of the neutrino mass matrix is assumed to have a pseudo-Dirac
form, to give maximal atmospheric neutrino mixing angle, it typically
gives $\Delta m^2_{atm} \ll \Delta m^2_{sol}$, which is wrong.

It is quite difficult to find a form of $M_{\nu}$ that both gives
maximal atmospheric neutrino mixing angle and $\Delta m^2_{sol} \ll
\Delta m^2_{atm}$. The ingenious model of Ref. \cite{mn} shows what is
required to obtain this result.

It is much easier to obtain maximal atmospheric neutrino mixing through
the charged lepton mass matrix \cite{hall},
precisely because that decouples the
neutrino mixing pattern from the neutrino mass pattern. Consequently,
the requirement that $\Delta m^2_{sol} \ll \Delta m^2_{atm}$ presents no
difficulty. All that is needed is that there be a nonabelian symmetry
relating $\mu^-$ and $\tau^-$ so that the parameter $\rho$ in Eq. (1)
comes out to have magnitude 1. This can be done in various ways. One
possibility is that $(\mu^-_L, \tau^-_L) \equiv \psi^-_i$ form a doublet of the
permutation group $S_3$, while the $e^-_L$ is a singlet. If the
$S_3$ is broken by a doublet ``flavon" field $\chi$, with its VEV given by
$\langle \chi_i
\rangle = (1,i)$ (this form of the VEV can emerge from certain simple
forms of the Higgs potential as shown below),
then the desired (3,2) and (3,3) elements of the matrix $L$
given in Eq. (1) can arise from the term $\tau^+_L \psi^-_i \langle \chi_i
\rangle \langle H_d \rangle$.  This will make $|\rho| = 1$ in Eq. (1).

Simple Higgs potentials can be constructed with flavon fields that have the
desired VEV structure.  As an example, consider the potential corresponding
to an $S_3$ doublet flavons $\chi \equiv (\chi_1, \chi_2)$.  The renormalizable
potential for $\chi$ that is invariant under $S_3$ as well as a $U(1)$ symmetry is
\cite{pakvasa}
\begin{eqnarray}
V(\chi) &=& \mu^2(\chi_1^*\chi_1+\chi_2^*\chi_2) + \lambda_(\chi_1^*\chi_1+\chi_2^*
\chi_2)^2+\lambda_2(\chi_1^*\chi_2-\chi_2^*\chi_1)^2 \nonumber \\
&+& \lambda_3[(\chi_1^*\chi_2+
\chi_2^*\chi_1)^2+(\chi_1^*\chi_1-\chi_2^*\chi_2)^2]~.
\end{eqnarray}

\noindent The VEVs can be parametrized as $\left\langle \chi_1 \right\rangle \equiv
r\cos\theta$, $\left\langle \chi_2 \right\rangle \equiv r\sin\theta e^{i\phi}$.
Minimization of $V$ with respect to $\phi$ and $\theta$ leads to $\phi = \pm \pi/2$
and $\theta = \pm \pi/4$, corresponding to $(\lambda_2+\lambda_3)$ having positive
sign.  This is the desired VEV, written as $\left\langle \chi \right\rangle \equiv
r(1,i)$.  Realistic charged fermion masses can be induced by making use of
flavon fileds which get VEVsof the form $(1,0)$ and $(0,1)$ in the space of the
second and the third families.  
For an analysis of alternative ways of inducing this VEV structure in the
context of supersymmetric models see Ref. \cite{hall}.  

If $|\rho| = 1$ in Eq. (1), the following relations obtain:
$\tan \theta_{sol} = \rho'/\rho = \rho'$, and $\tan \theta_{atm} =
\sigma = \sqrt{\rho^2 + \rho^{\prime 2}} = \sqrt{1 + \rho^{\prime 2}}$.
Together these imply that

\begin{equation}
\tan^2 \theta_{atm} = 1 + \tan^2 \theta_{sol},
\end{equation}

\noindent
or, for the more usually quoted quantity,

\begin{equation}
\sin^2 2 \theta_{atm} =
\frac{1 + \tan^2 \theta_{sol}}{(1 + \frac{1}{2} \tan^2 \theta_{sol})^2}.
\end{equation}

One sees that as $\tan^2 \theta_{sol}$
varies between 0 and 1, $\sin^2 2 \theta_{atm}$ varies between 1 and
$8/9$. (The point $\tan^2 \theta_{sol} = 1$, $\sin^2 2 \theta_{atm} =
8/9$, is the same as the prediction of flavor democracy models in the
pure flavor democracy limit.)  For the currently favored ``best fit'' value
of $\tan^2\theta_{sol} \simeq 0.4$, we have $\sin^22\theta_{atm} \simeq
0.97$, in excellent agreement with data.

\vspace{0.2cm}

\vspace{0.1in}
\noindent {\large \bf \S 4. Embedding in grand unified models}
\vspace*{0.1in}

One of the main virtues of lopsided mass matrices, which has been
emphasized in the literature \cite{bb, abb, syilr, ab2, lopsided}, is
that in the context of grand unified
theories they very elegantly account for the disparity between the
observed 2-3 mixings in the quark and lepton sectors, i.e. the fact that
$U_{\mu 3} \cong 0.7$ whereas $V_{cb} \cong 0.04$. The explanation
lies in the fact that $SU(5)$ relates the charged lepton mass matrix $L$
to the {\it transpose} of the down quark mass matrix $D$. In fact,
in the ``minimal" $SU(5)$ model $L = D^T$ exactly. In the form of
$L$ shown in Eq. (1), it is the $O(1)$ elements $\rho$ and $\rho'$ that
control the mixing of the left-handed fields and give large $U_{\mu 3}$.
However, if $D$ is
similar to the transpose of this form, then it is the small entry
$\epsilon$ (cf. Eq. (1)) that controls the mixing of the left-handed
down quarks of the second and third family, namely $V_{cb}$.
It should be noted that $SU(5)$ relates $L$ only to $D$, and not to
the up quark mass matrix $U$ or the neutrino Dirac mass matrix $N$.
Therefore, one expects that $D$ should be lopsided if $L$ is, but there
is no reason to suppose that $U$ and $N$ are. In fact, in lopsided
models that give a good account of quark and lepton masses and mixings,
only $D$ and $L$ are assumed to have lopsided forms. This is true also
of the realistic $SO(10)$ lopsided models that have been constructed
\cite{abb, ab2}.

Where the form in Eq. (1) differs from most published lopsided models
is in the large element $\rho'$. (However, the model of Ref. \cite{bb},
had a form for $L$ much like Eq. (1), with an entire row of large
elements.) The presence of the large element $\rho'$ puts significant
constraints on the building of realistic models of quark and lepton masses.
The point has to do with the so-called Georgi-Jarlskog factors of 3:
$m_s \approx m_{\mu}/3$ and $m_d \approx 3 m_e$ (at the GUT scale) \cite{gj}.
If the charged lepton mass matrix has the form shown in Eq. (1), then
the simplest way to get the first Georgi-Jarlskog factor is by assuming
that $D$ has the form:

\begin{equation}
D = \left( \begin{array}{ccc}
- & - & \rho' \\ - & - & \rho \\ - & -\epsilon/3 & 1 \end{array}
\right) m_D.
\end{equation}

\noindent
The factor of $-1/3$ in the $\epsilon$ term relative to the corresponding
term in $L$ is easily explained as being due to the $SO(10)$
generator $B-L$. Exactly this factor appears in the $SO(10)$ lopsided
models of Refs. \cite{abb, ab2}.

To get the second Georgi-Jarlskog factor of three requires, as is well-known,
that $\det D \cong \det L$. Barring some accidental cancellations, this
forces the (1,1) and (1,2) elements of $L$ (and correspondingly the (1,1) and (2,1)
elements of $D$) to vanish, or at least to be negligibly small. One can also,
without loss of generality rotate to make the (2,1) element of $L$ and the
(1,2) element of $D$ vanish. This leads to the virtually unique forms for $D$
and $L$. It is straightforward to generalize the $SO(10)$ model of
Refs. \cite{abb, ab2} to obtain the following realistic mass matrices:

\begin{equation}
\begin{array}{ll}
L = \left( \begin{array}{ccc} 0 & 0 & \delta' \\
0 & \delta & \epsilon \\
\rho' & \rho - \epsilon & 1 + \kappa \end{array} \right)m_D, \;\;\; &
D = \left( \begin{array}{ccc} 0 & 0 & \rho' \\
0 & \delta & \rho + \epsilon/3 \\ \delta' & -\epsilon/3 & 1
+ \kappa \end{array}
\right) m_D, \\ \\
N = \left( \begin{array}{ccc} 0 & 0 & 0 \\ 0 & 0 & \epsilon \\
0 & -\epsilon & 1 \end{array} \right) m_U, \;\;\; &
U = \left( \begin{array}{ccc} 0 & 0 & 0 \\ 0 & 0 & - \epsilon/3 \\
0 & \epsilon/3  & 1 \end{array} \right) m_U.
\end{array}
\end{equation}

\noindent
These mass matrices arise from the following Yukawa terms:
The entries denoted `1' come from $({\bf 16}_3 {\bf 16}_3) {\bf 10}_H$.
The $O(1)$ elements $\kappa$, $\rho$, and $\rho'$ come from
$({\bf 16}_3 {\bf 16}_H)({\bf 16}_i {\bf 16}'_H)$, where $i=1,2,3$,
the multiplets in the parentheses are contracted into $SO(10)$ vectors;
${\bf 16}_H$ breaks $SO(10)$ down to $SU(5)$,
and the ${\bf 16}'_H$ breaks the electroweak interactions. The
elements $\epsilon$, which appear antisymmetrically, come from
${\bf 16}_2 {\bf 16}_3 {\bf 10}_H {\bf 45}_H$, where the VEV of the
adjoint Higgs lies in the $B-L$ direction. Such an adjoint VEV is what
would be desired to achieve doublet--triplet splitting without
fine-tuning via the Dimopoulos-Wilczek mechanism in $SO(10)$ \cite{dw}.
The elements $\delta$ and $\delta'$
come from terms of the same form as the $\kappa$, $\rho$ and $\rho'$
terms, but with of course different family indices. These are exactly
the same kinds of operators that appear in the models of Refs.
\cite{ab1, abb, ab2}.

These mass matrices give a quite satisfactory fit to all
the quark and lepton masses and mixings.
In the approximation
$1\sim \rho \sim \rho' \sim \kappa \gg \epsilon \gg \delta \sim
\delta'$, so that the observed mass hierarchy is correctly reproduced,
the following mass relations are obtained at the GUT scale:
\begin{equation}
m_b \cong m_\tau,~~ m_s \cong m_\mu/3,~~
m_d \cong 3 m_e,~~m_u/m_t \cong 0~.
\end{equation}
The first three are the Georgi--Jarlskog relations, all of which work
quite well when compared with experimental values of the masses.  $m_u/m_t$ is
predicted to be zero by these forms. Experimentally, it is about
$10^{-5}$, which is about two orders of magnitude less than the corresponding
ratio for the down quarks, $m_d/m_b$. A tiny non-zero value of $m_u$ can
easily arise from some higher-dimension operator.

The parameter $\kappa$ in Eq. (10) is necessary in order to have adequate
CP violation in the CKM matrix.  We may redefine $1+\kappa$ to be simply
1 with an appropriate redefinition of $m_D$ in Eq. (10).  The parameter
$\epsilon$ in $L$ and $D$ of Eq. (10) will then be different from $\epsilon$
in $N$ and $U$.  Let us then rename $\epsilon$ appearing in $L$ and $D$
as $z\epsilon$.  In this redefined notation (we denote the redefined
$\delta, \delta',\rho, \rho'$ by the same symbols) we have $m_s/m_b \cong
\sigma/(1+\sigma^2)(z\epsilon/3), V_{us} \cong \delta'/(z\epsilon/3)$,
$V_{ub} \cong \delta'/(1+\sigma^2),~V_{cb} \cong (\epsilon/3)(z/(1+\sigma^2)-1)$.
From these relations, we obtain the following prediction:

\begin{equation}
\tan \overline{\theta}_{atm} = \frac{\tan 2 \theta_C (m_s/m_b)}{2 |V_{ub}|},
\end{equation}

\noindent
where $\theta_C$ is the Cabibbo angle, and
$\overline{\theta}_{atm}$ is the atmospheric neutrino
mixing angle that comes from the charged lepton matrix. (The contribution 
from the neutrino sector is assumed to be small.)  Note that 
$\overline{\theta}_{atm}$ is
of order unity, as needed for atmospheric neutrino oscillations.

The mixing parameter $U_{e3}$ is predicted to be
\begin{equation}
|U_{e3}| \simeq {\sin\theta_C \over 3}|U_{\mu 3}| \simeq (0.04-0.05)
\end{equation}
where the factor $\sin\theta_C/3$ arises from the small rotation needed
to complete the diagonalization of $L''$ of Eq. (3). (The factor 3 arises
because $m_\mu \cong 3 m_s$.)  This prediction
will provide a test of this class of models.

If the parameter $z$ were equal to 1 (which will be the case when the entry
$\kappa$ is absent in Eq. (10)), then there will be not enough CKM type CP violation
in this model, as all the mixing angles become approximately real.  This is true even when
we allow for the parameter $\delta$ to be complex, since there is a cancellation
between the up and the down quark contribution in the phase of the CKM matrix.
Allowing for $z \neq 1$ (or $\kappa \neq 0$)
leads to the desired CP violation, since $z$ is complex.  It is interesting to note
that if $z$ were equal to 1, the charm mass will be predicted to be
$m_c(m_c) \cong (1.1-1.2)$ GeV \cite{abb}.  Furthermore, the relation $|V_{ub}| \cong
(m_s/m_b)^2|V_{us}/V_{cb}|$ will follow, which is in good agreement with
experimental values.

\vspace{0.1in}
\noindent {\large \bf \S 5. New class of lopsided mass matrices from an $I_{3R}$ adjoint}
\vspace*{0.1in}

In the preceding example we made essential use of an $SO(10)$ adjoint VEV along the
$B-L$ direction.  Now we show that quite simple and predictive mass matrices
can be derived in $SO(10)$ if the VEV of the single adjoint present in the model points
along the $I_{3R}$ direction ($I_{3R}$ stands for the third component of
the right--handed isospin).  There is a simple and elegant realization of
the lopsidedness of $D$ and $L$ in this scheme.  It is worth noting that
a single adjoint scalar with its VEV along $I_{3R}$ direction can lead
to a natural doublet-triplet splitting, just as in the case of a single $B-L$ adjoint
\cite{barr}.  The $I_{3R}$ adjoint also has some advantages is suppressing
Higgsino-mediated proton decay in supersymmetric $SO(10)$ \cite{new}.

Consider the case where the $B-L$ adjoint that was involved in generating
the lopsided mass matrices of Eq. (10) is replaced by an $I_{3R}$ adjoint.
The mass matrices will then have the form

\begin{equation}
\begin{array}{ll}
L = \left( \begin{array}{ccc} 0 & 0 & \delta' \\
0 & \delta & -\epsilon' \\
\rho' & \rho - \epsilon & 1 \end{array} \right)m_D, \;\;\; &
D = \left( \begin{array}{ccc} 0 & 0 & \rho' \\
0 & \delta & \rho - \epsilon' \\ \delta' & -\epsilon & 1
\end{array}
\right) m_D, \\ \\
N = \left( \begin{array}{ccc} 0 & 0 & 0 \\ 0 & 0 & \epsilon' \\
0 & \epsilon & 1 \end{array} \right) m_U, \;\;\; &
U = \left( \begin{array}{ccc} 0 & 0 & 0 \\ 0 & 0 & \epsilon' \\
0 & \epsilon  & 1 \end{array} \right) m_U.
\end{array}
\end{equation}

As in the previous section, the `1' entries arise from ${\bf 16}_3 {\bf 16}_3 {\bf 10}_H$
coupling.  There are two crucial differences compared to Eq. (10).  The entry
resulting from ${\bf 16}_2{\bf 16}_3{\bf 10}_H{\bf 45}_H$ has now
two group contractions.  These two are denoted in Eq. (14) as
$\epsilon$ and $\epsilon'$.  These entries are proportional
to the $I_{3R}$ charge, so that they are equal in $D$ and $L$
(similarly in $N$ and $U$).  Lopsided nature arises from the
$\rho$ entry generated through ${\bf 16}_2 {\bf 16}_3 {\bf 16}_H {\bf 16'}_H$
coupling.  The parameters $\rho, \rho' $ are assumed to be much larger than $\epsilon,
\epsilon'$.  This model then predicts the following relatins:
\begin{equation}
m_b \cong m_\tau,~~
m_s \neq m_\mu,~~m_d m_sm_b \cong m_d m_\mu m_\tau~.
\end{equation}
The inequality for $m_s$ follows since $m_s/m_b \cong |\epsilon
\sigma/(1+\sigma^2)|$, while $m_\mu/m_\tau \cong |\epsilon'\sigma/(1+\sigma^2)|$,
where $\sigma \equiv \sqrt{\rho^2+\rho'^2}$.  Thus, although the entries
$\epsilon, \epsilon'$, proportional to $I_{3R}$ do not distinguish $L$ from
$D$, and the $\rho$-type entries also by themselves do not distinguish $L$ and
$D$ (these entries do not break $SU(5)$), a combination of the two leads to the
breaking of $m_\mu = m_s$ relation, as desired.  Unlike in Eq. (10), there is
sufficient CP violation in the CKM matrix in this model even without an entry
like $\kappa$ of Eq. (10).

Working in the approximation $1 \sim \rho \sim \rho' \gg \epsilon \sim
\epsilon' \gg \delta \sim \delta'$, we obtain the following relations
for the masses:
$m_b \cong m_\tau \cong \sqrt{1+\sigma^2}m_D$, $m_s/m_b \cong [(\sigma \epsilon+
\delta^* \rho/\sigma)|]/(1+
\sigma^2)$, $m_\mu/m_\tau \cong [|(\sigma \epsilon'+\delta^* \rho/\sigma)|
/(1+\sigma^2)]$,
$m_dm_sm_b =m_e m_\mu m_\tau$, $m_c/m_t \cong \epsilon \epsilon'$,
$m_u/m_t \cong 0$, all at the unification scale.  The CKM mixing angles are
given by $|V_{us}| \cong \delta'/(\epsilon+\delta \rho/\sigma^2)$,
$|V_{ub}| \cong \delta'/(1+\sigma^2)$, $|V_{cb}| \cong |\epsilon(2+\sigma^2)-\delta^*
\rho|/(1+\sigma^2)$.  Here all parameters have been made real by field
redefinitions, except $\delta$.  The rephasing invariant CP violation
parameter $\eta$ is given by $\eta \equiv {\rm Im}\{(V_{ub}V_{cs})/(V_{us}V_{cb})\}
\cong 2 \epsilon\rho{\rm Im}(\delta)/\{\sigma^2(1+\sigma^2) |V_{cb}|^2\}$.  
From these relations, we obtain the following prediction for the
atmospheric neutrino oscillations:
\begin{equation}
\tan\theta_{atm} \cong \sigma \cong (m_s/m_b)|V_{us}|/|V_{ub}|~,
\end{equation}
which is analogous to Eq. (12).  We also have a quantitative prediction for
$\tan\theta_{sol} \equiv \rho'/\rho$.  This can be seen by noting that
$\sigma$ is determined from Eq. (16), $\epsilon'$ from $m_\mu/m_\tau$,
$\epsilon$ from $m_c/m_t$, $\rho {\rm Im}(\delta)$ from $\eta$, $\rho {\rm Re}\delta$
from $m_s/m_b$, and $\delta'$ from $V_{ub}$.  The determinant relation
$m_em_\mu m_\tau = |\rho \rho' \delta|$ then fixes $\rho'/\rho$.

Consider the input parameters taking the following values.  $|V_{us}| \simeq 0.215,~
|V_{ub}| \simeq 0.0036,~|V_{cb}| \simeq 0.0037,~\eta \simeq 0.33$,
and $m_c(m_c) \simeq 1.35$ GeV, $m_t = 175$ GeV, all at the
weak scale, along with $m_s/m_b|_{GUT} \simeq 1/65$ (corresponding
to $m_s(1~GeV) \cong 130$ MeV).  
This gives the prediction $\tan\theta_{atm} \cong \sigma \simeq 1.034$,
$\tan\theta_{sol} \cong 0.43$ and $U_{e3} \simeq 0.06$, all of which
are in reasonably good agreement with current neutrino oscillation data.

In summary, we have presented simple realizations of bimaximal neutrino
mixing pattern, making use of lopsided mass matrices for the fermions.
This idea has a natural embedding in unified $SO(10)$ models.  We have
presented two different realizations within $SO(10)$, one making use
of the traditional $B-L$ adjoint VEV, and a new class of models  making
use of an $I_{3R}$ adjoint VEV.  We were also able to derive in a simple
way the near maximal mixing for atmospheric neutrino oscillation angle,
as given in Eq. (8).

\vspace*{0.1in}
\noindent{\bf Acknowledgments}
\vspace*{0.1in}

KSB wishes to thank the Theory Group at Bartol Research Institute
for the warm hospitality extended to him during a summer visit when
this work started.  The work of KSB is supported in part by DOE
Grant \# DE-FG03-98ER-41076, a grant from the Research Corporation
and by DOE Grant \# DE-FG02-01ER-45684.  The work of SMB is supported
in part by DOE Grant \# DE-FG02-91ER-40626.

\end{document}